\documentclass[11pt]{article}
\usepackage{amsmath,epsf,amssymb,latexsym,enumerate,cite,shadow,array,color}
\usepackage{slashed}
\usepackage{graphicx,wasysym}

\DeclareFontFamily{U}{rsf}{} \DeclareFontShape{U}{rsf}{m}{n}{
  <5> <6> rsfs5 <7> <8> <9> rsfs7 <10-> rsfs10}{}
\DeclareMathAlphabet\Scr{U}{rsf}{m}{n} \makeatletter
\@addtoreset{equation}{section} \makeatother

\def\be{\begin{equation}}
\def\ee{\end{equation}}
\def\ba{\begin{array}}
\def\ea{\end{array}}

\newcommand{\bea}{\begin{eqnarray}}
\newcommand{\eea}{\end{eqnarray}}

\def\K{K{\"a}hler}
\def\rmd{{\rm d}}

\usepackage{ifpdf}
\ifx\pdfoutput\undefined
   \pdffalse
\else
   \pdfoutput=1
   \pdftrue
  \usepackage[pdftex]{hyperref}
\def\u0{{\underline 0}}

\def\url{{\underline {r+\ell}}}

\newcommand{\rf}[1]{(\ref{#1})}
\newcommand{\vp}{\varphi}

\begin{document}

\begin{titlepage}

\begin{flushright}
CERN-PH-TH/2014-151
\end{flushright}

\hskip 1cm

\vskip 2cm

\begin{center}

{\LARGE \textbf{ Cosmology with  Nilpotent Superfields \\[0pt]
\vskip 0.8cm }}

\

{\bf Sergio Ferrara},$^{1,2,3}$\  {\bf Renata Kallosh}$^{4}$\   {\bf and  Andrei Linde}$^{4}$
\vskip 0.5cm
{\small\sl\noindent $^{1}$ Physics Department, Theory Unit, CERN, CH 1211, Geneva 23, Switzerland\\
$^{2}$ INFN - Laboratori Nazionali di Frascati, Via Enrico Fermi 40, I-00044 Frascati, Italy\\
$^{3}$ Department of Physics and Astronomy, University of California Los Angeles, CA 90095-1547 USA\\
$^{4}$ Department of Physics, Stanford University, Stanford, CA
94305 USA}
\end{center}
\vskip 1 cm

\begin{abstract}

We discuss N=1 supergravity inflationary models based on two chiral multiplets, the inflaton and the  goldstino superfield.  Using superconformal methods for these models, we propose to replace the unconstrained chiral goldstino multiplet by the nilpotent one associated with non-linearly realized supersymmetry of the Volkov-Akulov type. In the new cosmological models, the sgoldstino is proportional to a bilinear combination of  fermionic  goldstinos. It does not acquire any vev, does nor require stabilization, and does not affect the cosmological evolution. We explain a universal relation of these new models to $\kappa$-symmetric super-Dp-brane actions. This modification significantly simplifies a broad class of the presently existing inflationary models based on supergravity and string theory, including the simplest versions of chaotic inflation, the Starobinsky model, a broad class of cosmological attractors, the Higgs inflation, and much more. In particular, this is a step towards a fully supersymmetric version of the string theory axion monodromy inflation.  The new construction serves as a simple and manifestly supersymmetric uplifting tool in the KKLT-type string theory landscape. 
\end{abstract}

\vspace{24pt}
\end{titlepage}

\tableofcontents

\newpage

\section{Introduction}

There is a large class of general inflationary models  in  N=1 supergravity based on two chiral multiplets, the inflaton multiplet $\Phi$  and the   goldstino multiplet $S$. These  supergravity models,   where the chiral multiplet $S$ plays an important role during 
inflation, have the following \K\, potential and the superpotential 
\be
K= K(\Phi, \bar \Phi;  S, \bar S)\, , \qquad W= S\,f(\Phi)  \ .
\label{model}\ee
Here the inflaton superfield is 
\be
\Phi=  \phi+ i\,a + \sqrt 2\,  \theta\,  \chi + \theta^2 F^\Phi \ .
\ee
The inflaton $\vp$ can be either the  field $\phi$ or the field $a$, depending on which of these two fields is light during inflation.\footnote{Supergravity models without the $S$-multiplet typically have problems stabilizing one of these fields and keeping the other one light.  This is why we have sometimes referred to the $S$ field as a `stabilizer' field.} 
The goldstino  superfield is
  \be
S=s+ \sqrt{2}\, \theta\, G +\  \theta^2 F^S \ .
\label {s}\ee
Here $G$ is a goldstino fermion,  its supersymmetric scalar partner $s$ is  a sgoldstino, and $F_S$ is an auxiliary field of the goldstino multiplet. In 
 many of these models, the sgoldstino field vanishes during inflation, as well as after it,
\be
s=0\, .
\ee
For a partial list of such models see e.g. \cite{Cecotti:1987sa,Kallosh:2013lkr,Kawasaki:2000yn,Kallosh:2010xz,Kallosh:2014vja,Kallosh:2014xwa,Kallosh:2013pby,Cecotti:2014ipa,Kallosh:2014rga}.
Note that the goldstino direction is a direction in the moduli space where supersymmetry is broken and the corresponding auxiliary field does not vanish.  In our models during inflation  the auxiliary field in the inflaton direction vanishes, $F^\Phi= - e^{K/2} K^{\Phi \bar \Phi} \bar  W_{\bar \Phi} |_{s=0}=0$. The auxiliary field in the $S$ direction does not vanish,
$F^S= - e^{K/2} K^{S \bar S} \bar  W_{\bar S} |_{s=0}=- e^{K/2} K^{S \bar S} \bar f(\bar\Phi) \neq 0$. Therefore we refer to $S$ as a goldstino multiplet\footnote{The sgoldstino models of inflation\cite{AlvarezGaume:2010rt,Achucarro:2012hg} identify the inflaton with the scalar sgoldstino, a supersymmetric partner of the fermion goldstino, they are different from our models.} in models \rf{model}.

The first model in this class was constructed in the superconformal setting in \cite{Cecotti:1987sa}. It was shown in  \cite{Kallosh:2013lkr},  that it leads to supergravity version of the Starobinsky inflationary model  \cite{Starobinsky:1980te}  when supplemented by the stabilization terms in the \K\, potential of the form $(S\bar S)^2$. A supergravity model of a quadratic chaotic inflation  \cite{Linde:1983gd}
was proposed in \cite{Kawasaki:2000yn}, where the \K\, potential has a shift symmetry broken by the superpotential. A  large class of  supergravity models with shift symmetric \K\, potential leading to generic chaotic inflationary potentials  was found in \cite{Kallosh:2010xz}. Various recent examples of such models with shift symmetry broken by superpotential as well as by \K\, potential were presented in  \cite{Kallosh:2014vja,Kallosh:2014xwa}.  A different variety of these models, the so called `cosmological attractors'  \cite{Kallosh:2013pby,Cecotti:2014ipa,Kallosh:2014rga},  also belong to this class, they generalize the ones in  \cite{Cecotti:1987sa}. The superpotential is linear in $S$, however, the \K\, potential is not shift symmetric. 

In all our supergravity models  in\cite{Cecotti:1987sa,Kallosh:2013lkr,Kawasaki:2000yn,Kallosh:2010xz,Kallosh:2014vja,Kallosh:2014xwa,Kallosh:2013pby,Cecotti:2014ipa,Kallosh:2014rga}, there is one light scalar, the inflaton. The three other scalars are supposed to be very heavy so that they quickly vanish during inflation. The inflationary cosmology effectively becomes the single field inflation. The effective potential depends only on one inflaton scalar, $\vp$:
\be
V_{eff}(\vp) = e^{K(\Phi) }K^{S\bar S} |f(\Phi)|^2 \geq 0 \ .
\label{eff}\ee
In many of these models, it is relatively easy to achieve vanishing of the field orthogonal to the inflaton field $\vp$  \cite{Kallosh:2010xz}. However, in most of these models, one should take additional steps to stabilize the field $S$. Otherwise it either drifts from the minimum of the potential due to quantum fluctuations \cite{Demozzi:2010aj}, or becomes tachyonic, which leads to a major instability. Typically this problem can be cured by adding higher order stabilization terms, such as $(S\bar S)^{2}$, to the \K\ potential. While this procedure is legitimate, it makes the models more complicated, and it forces us to verify stability of each of such models, which is not always easy.

The purpose of this paper is to replace the unconstrained chiral goldstino superfield $S$ in eq. \rf{s}, which is a  cornerstone of all inflationary models in eq. \rf{model}, by the nilpotent superfield 
 \be 
 S^2(x, \theta)=0
 \, .
 \ee
The coupling of the nilpotent $S$ superfield to N=1 supergravity, in absence of other multiplets, just reproduces the result of \cite{Deser:1977uq}.
Such a  nilpotent superfield was proposed and  studied in the context of the Volkov-Akulov (VA) goldstino theory \cite{Volkov:1973ix} in 
\cite{rocek,Komargodski:2009rz,adgt,Kuzenko:2010ef}.  A replacement of this kind was already made in \cite{Antoniadis:2014oya} for the supergravity inflationary model based on \cite{Cecotti:1987sa}. It was shown there that the nilpotent  superfield $S$ leads to a VA type of an action coupled to the inflaton multiplet reproducing the Starobinsky potential. 
 Here we will introduce a nilpotent goldstino multiplet for generic inflationary models in eq. \rf{model} which were studied in \cite{Cecotti:1987sa,Kallosh:2013lkr,Kawasaki:2000yn,Kallosh:2010xz,Kallosh:2014vja,Kallosh:2014xwa,Kallosh:2013pby,Cecotti:2014ipa,Kallosh:2014rga}.
 
We will discuss here only  some basic results following \cite{Cecotti:1987sa,Kallosh:2013lkr,Kawasaki:2000yn,Kallosh:2010xz,Kallosh:2014vja,Kallosh:2014xwa,Kallosh:2013pby,Cecotti:2014ipa,Kallosh:2014rga}; their generalizations to other closely related models studied in the literature is straightforward. For example, we can build new models with a nilpotent superfield for Higgs inflation, starting with \cite{Ferrara:2010yw}. Many other models  studied e.g. in \cite{Yamaguchi:2000vm,Ellis:2014gxa} can be now modified and  used in the new construction with the nilpotent superfield.

 The immediate and obvious consequence of this step is  that the bosonic part of the inflationary models is simplified. Stabilization terms like $(S\bar S)^2$  vanish due to the nilpotent nature of the $S$ superfield. But these terms are also not required anymore in the new models since sgoldstino vev, the scalar component of goldstino is absent, being replaced by a bilinear of the fermions,  so there is no need to stabilize it. Therefore in these models one has to stabilize only the inflaton partner, one of the fields in the inflaton multiplet. In many of such models, this does not require additional stabilization terms \cite{Kallosh:2010xz}.\footnote{There is an alternative class of models suggested by $R+R^2$ supergravity in new minimal formulation \cite{Cecotti:1987qe} where the problem of moduli stabilization does not arise at all since the inflaton is the only scalar (member of a massive vector or tensor multiplet) but the \K\,  manifold in which it is embedded may change \cite{Ferrara:2013rsa}. These models, with a pure D-term potential, can interpolate between Starobinsky and chaotic inflation \cite{Kallosh:2014rga} for example, by changing the curvature of the $SU(1,1)/U(1)$   symmetric space. }
 
 The  significant consequence of involving the nilpotent goldstino chiral multiplet is a connection of the new versions of inflationary supergravity models in eq. \rf{model} supplemented with the $S^2=0$ requirement, to string theory.   Specifically, we are using the connection between the  super-Dp-branes  \cite{Cederwall:1996pv,Aganagic:1996nn,Bergshoeff:1997kr} and the Volkov-Akulov theory  \cite{Volkov:1973ix},    following \cite{Kallosh:1997aw,Bergshoeff:2013pia}. One of the important  early papers on the relation between the 3-brane actions, constrained superfields, and non-linear realization of supersymmetry is \cite{Rocek:1997hi}. A large list of references in that paper is relevant to our studies of new models of string cosmology
based on supergravity models where the unconstrained chiral goldstino multiplet $S$ has to be replaced by the nilpotent one, $S^2=0$.

As we are going to show in the paper, if we relate the nilpotent goldstino chiral multiplet to the previously studied models where we were able to stabilize the field $s$ at $s =0$, the bosonic part of the new class of the models emerges as a trivial generalization/simplification of the previously studied models. We take the potentials of the previously studied models obtained by the standard rules applied to unconstrained fields, and in the end simply take $s =0$. All previously obtained results describing inflation and its observational consequences in the models \cite{Cecotti:1987sa,Kallosh:2013lkr,Kawasaki:2000yn,Kallosh:2010xz,Kallosh:2014vja,Kallosh:2014xwa,Kallosh:2013pby,Cecotti:2014ipa,Kallosh:2014rga} remain intact.\footnote{The only difference appears in the models where the fluctuations of the field $s$ are interpreted as the curvaton perturbations  \cite{Demozzi:2010aj}; these perturbations are absent in the new scenario.}

On the other hand, it is {\it not}\, necessary to relate the new models to the previously studied ones with $s=0$. In extended versions of the models  \cite{Cecotti:1987sa,Kallosh:2013lkr,Kawasaki:2000yn,Kallosh:2010xz,Kallosh:2014vja,Kallosh:2014xwa,Kallosh:2013pby,Cecotti:2014ipa,Kallosh:2014rga} one may encounter many other moduli interacting with each other, which may lead to a cosmological time-dependent evolution of the field $s$. Successful chaotic inflation models of this type do exist, see e.g. \cite{Kallosh:2011qk}. However, the necessity to follow the cosmological evolution of all moduli and to ensure stability of an inflationary trajectory makes these models much more complicated to study. In essence, one should study everything numerically, and repeat it many times for different parameters to fully understand the dynamical features of the model.

In this respect,  models with a nilpotent goldstino chiral multiplet (or many such multiplets) provide additional advantages. One may derive potentials of such models using the standard rules, as if all fields there were unconstrained, and then, instead of investigation of the evolution of some of the fields, one may simply declare that they are nilpotent and therefore vanish. If this is a consistent approach, as we will argue in this paper, the theory is immediately simplified. If the original theory predicted that the field $s$ did not vanish, the predictions of the new theory will differ from the predictions of its non-constrained counterpart. However, the predictions of the new theories are much easier to study. We expect that this may stimulate development of many new inflationary models, which previously have been hampered by the necessity to control too many moduli simultaneously.

An interesting situation emerges in the new version of the supersymmetric Higgs inflation, when we use the model constructed in \cite{Ferrara:2010yw} as a starting point, where is corresponds to the NMSSM. The gauge singlet $S$ is an extra superfield which makes all the difference between the NMSSM and MSSM. Meanwhile, the same model which we constructed in \cite{Ferrara:2010yw} modified to involve a nilpotent superfield $S$, is neither NMSSM nor MSSM, since is has a non-linearly realized supersymmetry due to the nilpotent chiral superfield.
The cosmological properties of this model are the same as in \cite{Ferrara:2010yw}. However, the fermionic part of the action is new and has a Volkov-Akulov fermion without a scalar partner. It would be interesting to explore phenomenological implications in particle physics of this generalization of the supersymmetric standard model.

Another problem studied in this paper is to construct, using the nilpotent chiral multiplets, a new mechanism of uplifting the vacua in the stringy landscape. We will show that the updated O'KKLT models \cite{Kallosh:2006dv}  combining KKLT-type constructions \cite{Kachru:2003aw,Kallosh:2004yh}  with a nilpotent scalar multiplet $S$
\be
W=W_{\rm KKLT} (\rho)- \mu^2 S\, , \qquad K= - 3 \ln(\rho + \overline{\rho})+ S\bar S \qquad {\rm at} \qquad S^2=0 \ .
\label{KKLT 1}\ee
provides a manifestly supersymmetric uplifting of AdS vacua to de Sitter one, the fermionic part of the action being  of a VA-type. Our notation above, where we present some $S$-dependent $W$ and $K$ with the note `at' $S^2=0$ means that first the bosonic part of the action has to be computed, treating $S$ as a standard chiral superfield, and only at the end of the computations of the bosonic action one has to take it at $s=\bar s=0$, respecting the fact that we have imposed the operator relation $S^2=0$.

In essence, previously, in \cite{Kallosh:2006dv},  we were adding the Polonyi field \cite{Polo}  with the superpotential $\mu^2 S$, which provided the F-term uplifting of the KKLT models.  Investigation of the original versions of these models and their various generalizations often was complicated and required numerical analysis, in part because the value of the field $s$ after the uplifting no longer vanished. Meanwhile, in the new class of models, everything becomes nearly trivial: The term $- \mu^2 S$ provides a positive term in the potential, which no longer depends on $s$, as if the field $s$ in the original versions of these models were infinitely strongly stabilized at $s = 0$. This leads to enormous simplification of the F-term uplifting in string theory, and provides its string theory interpretation by using the relation between the  super-Dp-branes and the Volkov-Akulov theory.

To develop a fully consistent theory based on this mechanism, we would also need to analyze the fermionic part of the action. In this paper, we considered only its bosonic part which is necessary for investigation of inflation and vacuum stabilization.

 In this paper we will discuss only the F-term cosmological models in supergravities supplemented by the nilpotent chiral multiplet. Meanwhile there are well-known inflationary models in supergravity associated with the D-term potential \cite{Binetruy:1996xj}, \cite{Buchmuller:2013uta}. 
Also D-term  uplifting models \cite{Burgess:2003ic} may be studied  in the new constructions. For all these D-term models the update towards using the nilpotent superfields has to be studied separately.

 The paper is organized as follows. In sec.\,2 we use the superconformal version of supergravity and present a manifestly supersymmetric model with chiral superfields. Some of these superfields are  satisfying algebraic constraints when equation of motion for the superfield Lagrange multipliers is satisfied. In particular, these are models with nilpotent chiral superfields. In sec.\,23 we review the known facts about the relation between Dp-branes and VA goldstino action, and we explain that the fields of d=10 supergravity interact with VA goldstinos, which are fermions living on the world-volume of the Dp-brane. In sec.\,4 we discuss new cosmological models with a nilpotent chiral multiplet and their generic relation to string theory. In sec.\,5 we study the manifestly supersymmetric KKLT type uplifting with a non-linearly realized supersymmetry and with VA type fermions. We summarize the results in sec.\,6 and point our that more investigations will be necessary to relate specific string theory models to  d=4 N=1 supergravity with the nilpotent superfields.

\section{Superconformal Models Underlying Supergravity with  Nilpotent Chiral Multiplets}
We start with  superconformal model underlying N=1 supergravity interacting with some number of chiral multiplets $X^I$, $I=0,..., n$, in the form used for cosmological applications in \cite{Kallosh:2000ve}, based on \cite{cfgvp}. These models were further developed in  \cite{Ferrara:2010yw} with the emphasis on the Jordan frame and transition to Einstein frame supergravity,  while  building Higgs inflation and NMSSM inflationary models. 

 These theories  are described in details in \cite{SUGRA} starting with the action in eq. (17.15) there. Here we will  consider models without vector multiples in which case the superconformal action in eq. (17.15) has two terms
\be
[{\cal N}(X, \bar X)]_D + [{\cal W}(X)]_F\, , 
\label{conf}\ee
where ${\cal N}(X, \bar X)$ is a generic \K\, manifold potential of the embedding space, including the compensator superfield, it has a Weyl weight 2.  ${\cal W}(X)$ is the superpotential of the Weyl weight 3. The subscripts $D$ and $F$ refer the extraction of the $D$ and $F$ terms in the corresponding superfield actions. The F-term potential in the superconformal model originate from the auxiliary fields of the chiral multiplets
\be
V= -F^I \, {\cal N}_{I \bar J} \bar  F^{\bar J} - ( {\cal W}_I F^I + hc) \, , \qquad \Rightarrow \qquad V= {\cal W}_I   {\cal N}^{I \bar J} \bar {\cal W}_{\bar J} \ ,
\label{poten}\ee
since on-shell the value of the auxiliary fields of the chiral multiplets
is defined by the derivatives of the superpotential.
\be
F^I= -{\cal N}^{I \bar J} \bar {\cal W}_{\bar J} \ .
\ee
If we want to modify the theory \rf{conf} by making some of the chiral superfields to satisfy  algebraic constraints of the form $A_k(X)=0$, we can do it by adding a term with the chiral Lagrange multiplier\footnote{This technics of using superfield Lagrange  multipliers to algebraic constraints in the superconformal action was used from the early days of supergravity, for example in \cite{Cecotti:1987sa}. More recently it was used extensively in \cite{Kallosh:2013lkr} and in \cite{Cecotti:2014ipa}. } superfields
$\Lambda^k$ of Weyl weight 1 
 \be
[{\cal N}(X, \bar X)]_D + [{\cal W}(X)]_F + [\Lambda^k \, A_k(X)]_F\, ,
\label{confConstrA}\ee
where the functions $A_k(X)$ has to have a Weyl weight 2. The equation of motion over each superfield $\Lambda^k$ leads to our  set of algebraic superfield constraints 
\be
A_k(X)=0 \ ,
\label{constr}\ee
 since the Lagrange multipliers $\Lambda^k$  are present only in the $F$-terms. In particular, it is easy to add a constraint that one of the superfields, for example  $X^n$,  is nilpotent,   by adding the term $[\Lambda \, (X^n)^2]_F$ to the superconformal action. A detailed form of the supergravity action for chiral multiplets in presence of one or more  nilpotent ones will be presented in details in the future work, including the fermion part. Here we would like to stress that the action \rf{confConstrA} before the constraints \rf{constr} are solved, is manifestly supersymmetric.

Consider, for example,  a class of superconformal models useful for cosmology,  as described in \cite{Kallosh:2014ona}.  The  chiral multiplets in this case $X^I$, include the compensator field $X^{(0)}$, the inflaton  $X^{(1)}/X^{(0)}=\Phi$ and a goldstino   superfield $X^{(2)}/X^{(0)}=S$
\be
X^{(0)}, \qquad X^{(1)}= \Phi X^{(0)}, \qquad X^{(2)}= S X^{(0)} \ .
\ee
If we would like to replace the superfield $X^{(2)}$ by a nilpotent one in our general class of models, we have to start with the following
 superconformal    action
\be
[{\cal N}(X, \bar X)]_D + [{\cal W}(X)]_F + [\Lambda ({X^{(2)}})^2 ]_F \ .
\label{confConstr}\ee
{\it This action, before the constraint is solved, is manifestly supersymmetric.} All superfields including $X^{(0)}, X^{(1)}, X^{(2)}, \Lambda$ are standard unconstrained chiral multiplets of the conformal weight 1.
The equation of motion over $\Lambda$ leads to  algebraic superfield constraint $(X^{(2)})^2(x, \theta)=0$. The components of the unconstrained superconformal superfield 
\be
X^{(2)}(x, \theta)= x^{(2)}+ \sqrt 2 \theta \Psi^{(2)} + \theta^2 F^{(2)}
\ee
have to satisfy certain conditions \cite{rocek} and if $F^{(2)} \neq 0$ the constrained superfield becomes equal to
\begin{equation}
X^{(2)}|_{(X^{(2)})^2=0} \ = \ \frac{\Psi^{(2)}\Psi^{(2)}}{2 \, F^{(2)}} \ + \ \sqrt{2}\, \theta\, \Psi^{(2)} \ +\  \theta^2 F^{(2)}  \label{va12} \ .
\end{equation}
It means that the first component is a bilinear of the fermions.

Now we would like to proceed from the superconformal theory \rf{confConstrA} or \rf{confConstr} to supergravity.
In the superconformal gauge  where the compensator is fixed to be $X^0=\bar X^0 = \sqrt 3 M_P$ the supergravity model in the Jordan frame is recovered so that the first term in the action in \rf{conf} has a term
\be
  { 1 \over 2} \Omega (z, \bar z)\, R \ ,
\label{Jordan} \ee
where ${\cal N} (X,\bar X)|_{X^0=\bar X^0 = \sqrt 3 M_P}= -3 \, \Omega (z, \bar z)$ where $z^I= X^I/X^0= (1, z^i)$ which defines the  superfields  $z^i$. The superpotential and a potential of the Einstein frame supergravity for each model are deduced from ${\cal N}(X, \bar X)$ and  $ {\cal W}(X)$ of the superconformal model. In particular, the frame function defines the \K\, potential 
\be
 K(z, \bar z)= -3 \log \, \Omega (z, \bar z) = -3 \log \Big (- {1\over 3} \,  {\cal N} (X,\bar X)|_{X^0=\bar X^0 = \sqrt 3 M_P} \Big) \ .
\ee
It also means that after gauge-fixing $S= X^{(2)}/ \sqrt 3 M_P$. The nilpotent constraint on $X^{(2)}$ transfers on a nilpotent constraint on $S$, i. e. we have the 
condition
$
S^2(x, \theta)=0 \label{va0} 
$.
This eliminates the sgoldstino $s$ in favor of a goldstino bilinear in \rf{s}, so that 
\begin{equation}
S \ = \ \frac{GG}{2 \, F^S} \ + \ \sqrt{2}\, \theta\, G \ +\  \theta^2 F^S  \label{va1} \ .
\end{equation}
For example, the $\theta^2$ component of $S^2(x, \theta) $ following from \rf{s} is 
$2 s F^S - 4 G G $, and it must vanish. It follows that $s={GG\over 2 \, F^S}$ as we show in \rf{va1}. If  the auxiliary field $F^S$ is not vanishing one finds that sgoldstino is replaced by a fermion bilinear 
\be
s= \frac{\sum_{\alpha =1}^{\alpha=2}G^\alpha G_\alpha}{2 \, F_S}= \frac{\psi^1\psi^2}{ \, F_S} \ ,
\ee
where we used notation $G^1=\psi^1$ and $G^2= G_1= \psi^2$. The vev of $s$ must vanish since its square is given by an expression 
\be
s^2= {(\psi^1\psi^2)^2\over  (F^S)^2}=0 \ .
\ee
It  vanishes since each of the Grassmann variable  products vanishes,  $(\psi^1)^2= (\psi^2)^2=0$. Therefore the fermion bilinear replacing sgoldstino in our class of models  cannot  acquire a non-trivial vev, as different from examples in superconductivity or technicolor models where the fermion  bilinears form condensates and effectively replace scalars.

In the supergravity version of the superconformal theory 
$
F^S = -e^{K\over 2}  K^{S\bar S} \partial_{\bar S} \bar W 
$ which also is required to be not vanishing for the operator constraint $S^2=0 $ to be valid, so that 
\be
s \qquad \Rightarrow \qquad \frac{GG}{2 \, F^S} \ .
\ee
{\it The new superconformal action \rf {confConstr} means that the bosonic terms in the supergravity action with the nilpotent goldstino are the same as in models with unconstrained $S$  but taken directly at $s=0$.}  One has to keep the $S$ contribution in the \K\, potential and in the superpotential $W$  till the complete action is computed, according to standard rules. At the end of the computation, in the action one should set $s=0$ , to get the complete bosonic action in the model with the nilpotent multiplet $S$.

In other words, one can take all previously studied models of the type discussed above, including the models where the field $S$ was not stabilized, or where it was non-zero, or even time-dependent during the cosmological evolution. Then one should simply declare that $s=0$ in these models. If the field $s$ vanished in the original models, then the results obtained in the original models will coincide with the corresponding results in the new models with the nilpotent multiplet $S$. But if the field $s$ did not vanish in the original models, such as \cite{Kallosh:2011qk}, we should simply declare that $s = 0$ in those models, and repeat the rest of the investigation, which becomes a much simpler task.

The previous comments are valid for the investigation of the bosonic part of the models, which is typically sufficient to study inflation, or to investigate stability of the string theory vacua. Meanwhile {\it the fermionic part is significantly different from the standard action with unconstrained multiplets.} The reason for such complications is due to complicated equations of motion for auxiliary fields. In models without nilpotent superfields the action has quadratic and linear terms only, as shown in eq. \rf{poten}. Now some of the dependence in the fermionic part has terms $\frac{GG}{2 \, F^S}$, the procedure of solving for $F^S$ in fermionic part of the action becomes very complicated.
For example the former kinetic term for $s$ is replaced by a complicated function of fermions
\be
 s \, \partial ^2 \bar s  \qquad \Rightarrow \qquad   \frac{GG}{2 \, F^S} \partial ^2 \frac{\bar G \bar G}{2 \, \bar F^S}
\ee
and these and other terms contribute to equations of motion for auxiliary fields.

The simplest  $S$ superfield  supergravity model  \cite{Antoniadis:2014oya}  in absence of all other chiral fields is based on 
\begin{equation}
K \ = \ - \  \, \log\left(1 \,+\, {1\over 2}(S - \overline{S})^2\right) \ \equiv \ \, S \, \overline{S} \ , \qquad W \ = \ f\, S \  \ ,  \qquad {\rm at} \qquad S^2=0 \ ,
\label{va3}
\end{equation}
and leads to a potential
\be
V \ = f ^2  \ .
\label{pot}\ee
A non-gravitational part of this action,  in the form given in   \cite{Komargodski:2009rz} is
\be
{\cal L}_{VA}= - f^2 +i \partial_\mu \bar G \bar \sigma ^\mu G + {1\over 4 f^2} \bar G^2 \partial^2 G^2 - {1\over 16 f^6} G^2 \bar G^2 \partial^2 G^2 \partial^2 \bar G^2 \ ,
\label{VA}\ee
corresponding to a superfield action
\begin{equation}
{\cal L}_{VA} \ = \ \bigg[ S\, \overline{S} \bigg]_D \ + \ \bigg[f S \,+\Lambda S^2 + \, h.c. \bigg]_F \ . \label{va2}
\end{equation}
where $\Lambda$ is a Lagrange multiplier chiral superfield. It agrees with the original VA action \cite{Volkov:1973ix}, according to \cite{Kuzenko:2010ef}
after a spinorial field redefinition.
\section{ Review of the Super-D-branes and Dirac-Born-Infeld-Volkov-Akulov actions}

Our review of the super-Dp-branes and derivation of the DBI-VA actions from D-branes is based on \cite{Kallosh:1997aw} and mostly on the recent detailed studies in  \cite{Bergshoeff:2013pia} and references therein. Other aspects of relation of branes to supersymmetry breaking at the string scale were studied in \cite{Antoniadis:1999xk}.

Classical D-brane actions \cite{Cederwall:1996pv,Aganagic:1996nn} in supersymmetric string theory have a fermionic local symmetry, $\kappa$-symmetry. This local symmetry of the  classical D-brane actions has been gauge-fixed in the flat background in \cite{Aganagic:1996nn}. In the bosonic
 d=10 on-shell supergravity background the gauge-fixing of a local $\kappa$-symmetry was performed in \cite{Bergshoeff:1997kr}. 
The bosonic supergravity background, $G, B$ and $\phi$ includes the
spacetime metric, the NS/NS 2-form gauge potential and the dilaton,
respectively, as well as RR forms $C^{(r)}$ where $r=0,...,10$.  The bosonic Dp-brane is described by a map $X$ from
the worldvolume $\Sigma_{(p+1)}$ into the $d=10$ spacetime ${\cal M}$
and by a 2-form Born-Infeld field strength $F$ on $\Sigma_{(p+1)}$; $dF=0$ so
$F=dV$ where $V$ is the one-form Born Infeld  gauge potential. The bosonic part
of the effective action of a Dp-brane using notation of \cite{Bergshoeff:1997kr} is
\begin{equation} 
\label{action}
I_{\rm p} = -\int d^{p+1}\sigma\, \left[
e^{-\phi}\sqrt {|{\rm det} (g_{ij}+{\cal F}_{ij})|} + Ce^{{\cal F}} +
mI_{{\rm CS}}\right]\, ,
\end{equation}

\noindent where
$
g_{ij} = \partial_i X^\mu \partial_j X^\nu
G_{\mu\nu}\, 
$
 is the metric on $\Sigma_{(p+1)}$ induced by the map $X$,
$(\mu, \nu=0, \dots 9)$ are the spacetime indices and ${\cal F}_{ij}\ 
(i=1,\cdots (p+1))$ is the modified 2-form field strength 
\begin{equation}
\label{calF}
{\cal F} =  F - B\, ,
\end{equation}
\noindent where $F_{ij}$ is the Born-Infeld  2-form field strength 
and $B_{ij}$ in ${\cal F}_{ij}$ is the pull-back $B_{ij} = \partial_i X^\mu \partial_j X^\nu
B_{\mu\nu}\, $ of the NS-NS 2-form
gauge potential $B_{\mu\nu}$ with $X$. The second term in (\ref{action}) is a
Wess-Zumino-Chern-Simons term, where
\begin{equation}
C = \sum_{r=0}^{10} C^{(r)}
\end{equation}

\noindent is a formal sum of the RR gauge potentials $C^{(r)}$. 
It is understood that after expanding the potential only the
(p+1)-form is retained.  The last term in (\ref{action}) is only present for even p (the
IIA case). Its coefficient $m$ is the cosmological
constant of massive IIA supergravity and $I_{{\rm CS}}$.

For the supersymmetric Dp-brane actions,  the maps $X$
($\{X^\mu\}$) are replaced with supermaps $Z=(X,\theta)$ ($\{Z^M\}$) and the
various bosonic supergravity fields with the corresponding superfields
of which they are the leading component in a $\theta$-expansion.  

At this point things become rather technical in  \cite{Bergshoeff:1997kr}, however, there is one nice and simple feature in this construction: it is `democratic' in the sense that both type IIA as well as type IIB Dp-branes are described, with even and odd $p$, in the same construction. This is close in spirit to a    `democratic' version of d=10 supergravity  in \cite{Bergshoeff:2001pv}. The specific notation allows to
treat the IIA and IIB theories in a unified way.   The induced metric for both IIA and IIB D-branes
is given by the super-vielbeins
\begin{equation}
g_{ij} = E_i{}^aE_j{}^b\eta_{ab}\, ,\qquad 
E_i{}^A = \partial_i
Z^M E_M{}^A\, .
\end{equation}

We will now only explain the steps in deriving the Dirac-Born-Infeld-Volkov-Akulov actions from the generic super-Dp-branes which are relevant for our purpose here. Our purpose is to explain the relation between inflationary models of N=1 supergravity d=4 with the inflaton and a nilpotent goldstino multiplet and
super-Dp-branes. 

1. We will explain, following \cite{Kallosh:1997aw}, \cite{Bergshoeff:2013pia} why the goldstino action of the Volkov-Akulov type is part of the gauge-fixed supersymmetric Dp-brane actions. We will use the case of type IIB models for simplicity.

2. We will show that the fermionic goldstino interacts with the NS-NS 2-form $B_{\mu\nu}$ of the supergravity in d=10 as well as with other fields of the d=10 supergravity, including RR forms.

\subsection{ Dp-superbrane with local \texorpdfstring{$\kappa$}{kappa}-symmetry in the flat supergravity background }
\label{ss:Dpmax}

The $\kappa$-symmetric Dp-brane action in type IIB (with $p=2n+1$ odd), in a flat
background geometry with  coordinates
$X^{m}$, $m = 0,\dots , 9$, consists of the Dirac-Born-Infeld-Nambu-Goto
term $S_{\rm DBI}$ and Wess-Zumino term $S_{\rm WZ}$ in the world-volume
coordinates $\sigma^{\mu}$ $(\mu =0,\dots,p)$:
\begin{equation}
\label{actiongeneral} S_{\rm DBI} +S_{\rm WZ} =  -\frac{1}{\alpha^2}
\int \rmd^{p+1} \sigma\, \sqrt{- \det (G_{\mu\nu} + \alpha
{\cal F}_{\mu\nu})} +\frac{1}{\alpha^2}\int \Omega_{p+1} \,.
\end{equation}
Here $G_{\mu\nu}$ is the  manifestly supersymmetric induced world-volume
metric
\begin{equation}
G_{\mu\nu} = \eta_{mn} \Pi_\mu^m \Pi_\nu^n \ , \qquad \Pi_\mu^m =
\partial_\mu X^m - \bar\theta \Gamma^m \partial_\mu \theta \,,
\end{equation}
and the Born-Infeld field strength ${\cal F}_{\mu\nu}$ is
given by
\begin{equation}
{\cal F}_{\mu\nu} \equiv F_{\mu\nu} - b_{\mu\nu} \,, \qquad
b_{\mu\nu} = \alpha^{-1} \bar{\theta} \sigma_3
\Gamma_{m}\partial_{\mu}\theta\left(\partial_{\nu} X^{m}
-\frac{1}{2} \bar{\theta}\Gamma^{m}\partial_{\nu}\theta\right)-\left(
\mu\leftrightarrow \nu \right)\, ,
\label{defcalFbrane}
\end{equation}
where $\Omega_{p+1}$ is a particular $p+1$-form. Note that the superspace coordinates $Z(\sigma)= \Big (X(\sigma), \theta(\sigma)\Big )$ depend on the world volume coordinates $\sigma$. We use here notation of \cite{Bergshoeff:2013pia}.

The action has the global ($\sigma$-independent) supersymmetry on the world-volume of the brane
\begin{eqnarray}
\label{susytrans} \delta_\epsilon \theta &=& \epsilon\, , \qquad
\delta_\epsilon X^m = \bar\epsilon\Gamma^m\theta\, , \qquad
\nonumber\\ [.2truecm] \delta_\epsilon A_\mu &=&\alpha^{-1} \bar\epsilon
\sigma_3 \Gamma_m \theta
\partial_\mu X^m - \tfrac{\alpha^{-1}}{6} \left(\bar\epsilon \sigma_3 \Gamma_m \theta \bar\theta \Gamma^m
\partial_\mu\theta + \bar\epsilon \Gamma_m \theta\bar\theta \sigma_3 \Gamma^m \partial_\mu\theta\right)\,.
\end{eqnarray}
Besides the global supersymmetry the action is also invariant under a local ($\sigma$-dependent) $\kappa$-symmetry.
One can  gauge fix  $\kappa$-symmetry  and general
coordinate transformations in a covariant gauge  
discovered in \cite{Aganagic:1996nn}. The fermionic gauge in IIB models is of the form
\begin{eqnarray}
\theta_1(\sigma) &=& 0 \, ,   \qquad \theta_2 (\sigma)\equiv \alpha \lambda (\sigma)\ ,
\label{thetasB}
\end{eqnarray}
where $\theta^1(\sigma), \theta^2(\sigma)$ are two positive chirality spinors of type IIB theory, which are both functions of the world volume coordinates $\sigma$. The basic role of the gauge-fixing $\kappa$-symmetry is to control the correct number of degrees of freedom on the brane. The quantization allows to remove the half of fermionic fields $\theta_1(\sigma) $ from the brane action, the remaining half of the   fermion fields on the brane  $\theta_2 (\sigma)$ become the Volkov-Akulov type  goldstino's $\lambda (\sigma)$. In both type IIA and type IIB
models the WZ term vanishes in the flat background.  The gauge-fixed action of the Dp-brane at $\alpha=1$ has the form \cite{Aganagic:1996nn}
\be
S^{(p)}= -\int d^{p+1} \sigma \sqrt {-\det M^{(p)}} \ ,
\ee
as shown in eq. (85) in \cite{Aganagic:1996nn}, where the details can be found. For general $\alpha$ the derivation was given in  \cite{Bergshoeff:2013pia} \footnote{ In \cite{Bergshoeff:2013pia}  the Wess-Zumino term $\Omega_{p+1}$ was taken to be constant
 since we were only interested in the actions for spinors and vectors. However, now we are paying attention to the fact that the DBI part of the classical action survives the gauge-fixing whereas the WZ term vanishes in the gauge \rf{thetasB} in absence of the bosonic d=10 background, which results in the action given in \rf{actionBI} in agreement with \cite{Aganagic:1996nn}.
}.
For example for the D-9-brane, the gauge-fixed action is given by the Dirac-Born-Infeld-Volkov-Akulov action
\begin{equation}
\label{actionBI} S_{DBI-VA} =  -\frac{1}{\alpha^2} \int \rmd^{10} \sigma\,\left\{
\sqrt{- \det (G_{\mu\nu} + \alpha {\cal F}_{\mu\nu})}  \right\}\,,
\end{equation}
where 
\begin{eqnarray}
G_{\mu\nu} = \eta_{mn} \Pi_\mu^m \Pi_\nu^n  \qquad  \Pi_\mu^{m} &=& \delta^{m}_{\mu}
-\alpha^2\bar\lambda \Gamma^{m}
\partial_\mu \lambda \, ,\end{eqnarray}
\begin{eqnarray}
{\cal F}_{\mu\nu} &\equiv& F_{\mu\nu} - 2\alpha\bar{\lambda}\Gamma_{[\nu }\partial_{\mu]}\lambda\, .
\end{eqnarray}
The d=4 counterpart of \rf{actionBI} for N=2 supersymmetry spontaneously broken down to N=1 is the N=1 manifestly supersymmetric Born-Infeld action \cite{Cecotti:1986gb}. It was shown 
 to have a second nonlinearly realized supersymmetry acting on the N=1 field strength superfield in  \cite{Bagger:1996wp}. A detailed study of related issues of partial breaking of global d = 4 supersymmetry, constrained superfields, and 3-brane actions was performed in
  \cite{Rocek:1997hi}.

The  formula in \rf{actionBI} at  $\alpha=1$ was first derived and presented in eq.  (1) in \cite{Aganagic:1996nn}. Meanwhile  in \cite{Kallosh:1997aw} it was observed that in absence of fermions $\lambda(\sigma)$ we recover the classical  supersymmetric DBI models, for
example in $d=10$ we find
\begin{equation}
\label{simpleBI}
S_{\rm DBI} =  -\frac{1}{\alpha^2} \int \rmd^{10} x\,\left\{ \sqrt{- \det (\eta_{\mu\nu} + \alpha  F_{\mu\nu})}\right\} \ .
\end{equation}
On the other hand,  when the covariant 2-form $
F_{\mu\nu}$ is absent, the same action is a $d=10$ analog of the $d=4$ VA action \cite{Volkov:1973ix}, as explained in \cite{Kallosh:1997aw}
\begin{equation}\label{simpleVA}
S_{\rm VA} =  - \frac{1}{\alpha^2} \int {\rm d}^{10} x\,
\sqrt{- \det G_{\mu\nu} } =  \frac{1}{\alpha^2} \int E^{m_0} \wedge ...\wedge E^{m_9} \ ,
\end{equation}
\begin{equation}\label{E}
E^m={\rm d} x^m+ \alpha^2 \bar\lambda \Gamma^{m}
{\rm d} \lambda \, .
\end{equation}
Note that our parameter $\alpha$ in \cite{Bergshoeff:2013pia} is inversely proportional to the parameter $f$ in the VA action in \rf{VA}.
In this truncated model the
exact hidden  non-linear supersymmetry transformation of fermions consists of two terms, one is a 
 shift, and the other one is an expression which is quadratic in fermions. This is  literally the
original Volkov-Akulov formula
\begin{equation}
\delta_{\zeta} \lambda = \alpha^{-1}\zeta +\alpha
\bar{\lambda}\Gamma^{\mu}\zeta \partial_{\mu}\lambda\,.
\end{equation}
It signals the spontaneous breaking of a non-linearly realized supersymmetry on the brane due to the presence of the constant term $ \alpha^{-1}\zeta$ in the transformation rules, $\alpha$ being some finite constant. 

We have used here an example of D9 super-brane, as the simplest case of appearance of the VA goldstino's action, \cite{Kallosh:1997aw}. Meanwhile, as shown in \cite{Bergshoeff:2013pia}  this is a generic phenomenon for all super-Dp-branes as well as more exotic V-branes, discussed there. Note that when all fields, spinors and vectors, are absent, all these gauge-fixed Dp-brane actions  are equal to
\begin{equation}
\label{actionBIvac} S_{vac} =  -\frac{1}{\alpha^2} \int \rmd^{p+1} \sigma\,\left\{
\sqrt{- \det \eta_{\mu\nu} }  \right\}\,,
\end{equation}
and have positive energy density $f^2$ in agreement with the VA action \rf{VA} since $f^2= \alpha^{-2}$. In  our effective supergravity actions in \rf{model} and \rf{pot} we see an analogous contributions to the potential energy.

\subsection{Goldstino's interaction with  NS-NS 2-form $B$ and RR forms $C^{(r)}$ and the axion potential}
We now return to the super-Dp-brane action \rf{action} in the supergravity background. It means that  in the DBI-VA action there are
  terms like
\be
\sqrt{- \det (G_{\mu\nu} + \alpha {\cal F}_{\mu\nu})}\, ,   \qquad  \,  \,  \sum_{r=0}^{r=10} C^{(r)} \, e^{{\cal F}}\, .
\label{int}\ee
Upon gauge-fixing $\kappa$-symmetry these terms depend on the following combination
\begin{eqnarray}
{\cal F}_{\mu\nu} &\equiv& F_{\mu\nu} - 2\alpha\bar{\lambda}\Gamma_{[\nu }\partial_{\mu]}\lambda - B_{\mu\nu} +...\,
\end{eqnarray}
Since there are terms in the action with the non-linear dependence on ${\cal F}_{\mu\nu}$, there is an interaction between the bilinears of goldstino and the NS-NS 2-form field $B_{\mu\nu}$ of d=10 supergravity/string theory.

When the d=10 string theory with super-Dp-branes is compactified and studied in the form of N=1 supergravity, one may  associate the models with the superpotential 
\be
W= S\, f(\Phi)\, ,\qquad S^2=0 \ , 
\label{sp}\ee
with string theory super-Dp-branes interacting with the supergravity background. The condition for this association is that the chiral superfield $S$ is nilpotent, $S^2=0$,  and corresponds to a Volkov-Akulov goldstino model, whereas
 $\bar \Phi -\Phi$ describes the axion $\int_{\Sigma_2} B$ interacting with goldstinos in a supersymmetric way. This same axion was  used in axion monodromy models \cite{McAllister:2014mpa}. 
 
 Another source of interaction between the goldstino multiplet and a d=10 supergravity background fields might show up via the \K\, potential where the following interaction becomes possible
 \be
 K= S\bar S \pm{1\over 2}  (\Phi \pm \bar \Phi)^2 (1+ \gamma S\bar S) \ .
 \label{gamma}\ee
 These kind of models were used in \cite{Kallosh:2010xz} and it was shown there that the non-vanishing $\gamma$-terms help to stabilize the partner of the inflaton.

Note also that in various super-Dp-branes interacting with the background supergravity the WZ term in \rf{int} suggest that the RR forms $C^{(r)}$ also interact with goldstino's and therefore an inflationary multiplet $\Phi$ does not have to be related to the NS-NS 2-form, but might also originate from some RR fields. The dependence of the function $f(\Phi)$ in the superpotential on their holomorphic argument $\Phi$, polynomial or exponential,  is model dependent. It may depend on the particular string theory setting, which has to be studied in the context of specific string theory models.

\section{New Cosmological Models with the Nilpotent Superfield $S$}
Here we give an upshot of cosmological applications of new inflationary models with
\be
K= K(\Phi, \bar \Phi; S\bar S)\, , \qquad W= S\,f(\Phi)\, , \qquad S^2=0\, , \qquad V(\Phi)= e^{K(\Phi)} K^{S\bar S} |f(\Phi)|^2 \geq 0 \ ,
\label{model1}\ee
where $S$ is a nilpotent superfield. The total bosonic action for all of these models is the one we would have in case of the unconstrained $S$ but taken at the value of $s=0$. Therefore the new bosonic action does not have a kinetic term for $s$ scalars and all $s$ terms in the potential should be put to zero. This step is not for free in the complete supergravity action. The fermionic part of the total supersymmetric action differs significantly from the standard N=1 supergravity interacting with unconstrained chiral multiplets. This feature of new models  takes off the burden of stabilizing  the complex scalar $s$ from the $S$ multiplet, which was not easy in the same models where $S$ was an unconstrained superfield. The only remaining concern is the stabilization of the inflaton partner in $\Phi|_{\theta=0}= \phi+i a$. One of these scalars must be heavy, the other is light.  This is not easy to achieve in models with a single superfield, see for example \cite{Ketov:2014qha} for a recent discussion. However, in our case, both with an unconstrained goldstino as well as a nilpotent goldstino, this problem has an easy solution in many inflationary models.

\subsection{Chaotic inflation in supergravity}

We will begin with the generic chaotic inflation models in supergravity with 
\be
K= -{(\Phi-\bar\Phi)^2\over 2}+S\bar S \ , \qquad  W = m S \Phi \  ,
\ee
\cite{Kawasaki:2000yn,Kallosh:2010xz}. Representing the scalar component of the superfield $\Phi$ as a sum of canonically normalized fields $(\phi+ i\,a)/\sqrt 2$, one finds that the field $\phi$ plays the role of the inflaton field with the simplest quadratic potential 
\be
V(\phi) = {m^{2}\over 2}\phi^{2}
\ee
and the mass squared of the fields $\phi$, $a$ and $s$ near the inflationary trajectory $a = s = 0$ during inflation is given by 
\be
m^{2}_{\phi} = m^{2}, \qquad m^{2}_{s} = m^{2}, \qquad m^{2}_{a} = 6H^{2}+ m^{2} \ .
\ee
The field $a$ is strongly stabilized at $a = 0$, but the field $s$  has the same mass as the inflaton field, so its quantum fluctuations are generated during inflation. Depending on the details of the theory, these perturbations later may either become irrelevant, or lead to abundant isocurvature perturbations, or to adiabatic perturbations via the curvaton mechanism \cite{Demozzi:2010aj}.

One may consider a different version of this scenario \cite{Kallosh:2010xz}, with 
\be
K= -3 \log \left[1+ {(\Phi-\bar\Phi)^2\over 6}-{S\bar S\over 3} \right] \ , \qquad  W = m S \Phi 
\ee
The potential of the field $\phi$ will remain the same and before, $V(\phi) = {m^{2}\over 2}\phi^{2}$, but the masses of the fields $a$ and $s$ will be different. Most importantly, the field $s$ during inflation will become tachyonic, which destroys the inflationary regime. Fortunately, one can stabilize the field $s$ and get rid of its fluctuations by adding a sufficiently large term $\sim (S\bar S)^{2}$ to the \K\, potential. However, this makes the model more complicated and less predictive.

In the new version of these models, with the nilpotent superfield $S$, this problem disappears. One just takes $s =0$; the field $a$ is stable in both versions of the model, and the potential remains equal to $V(\phi) = {m^{2}\over 2}\phi^{2}$, as in the simplest version of the chaotic inflation scenario \cite{Linde:1983gd}.

Similar result is true for a more general scenario with
\be
K  =K ((\Phi-\bar \Phi)^2, S\bar S)  \ , \qquad  W = S f(\Phi) \qquad {\rm at} \qquad S^2=0 \ .
\ee
where $f(\Phi)$ is a real holomorphic function. If $S$ is nilpotent, no stabilization of the field $s$ is required, the field $a$ typically does not need stabilization, though it can be provided \cite{Kallosh:2010xz}, and the inflationary potential is given by 
\be
V(\phi) = |f(\phi/\sqrt 2)|^{2} \ .
\ee
Since the restriction that $f(\Phi)$ is a real holomorphic function is very mild (it is satisfied by any function which can  be represented as a series with real coefficients), this class of theories can describe any desirable set of the observable parameters $n_{s}$ and $r$  \cite{Kallosh:2010xz}, without any need to add extra terms higher order in S to the \K\, potential.

\subsection{ Inflationary models with the nilpotent superfields  related to string theory}
Inflationary  supergravity models in \cite{Cecotti:1987sa,Kallosh:2013lkr,Kawasaki:2000yn,Kallosh:2010xz,Kallosh:2014vja,Kallosh:2014xwa,Kallosh:2013pby,Cecotti:2014ipa,Kallosh:2014rga} do not seem to have any 
obvious relation to string theory. However, once the goldstino chiral superfield in all these models is replaced by the nilpotent multiplet, all these models have a simple relation to super-Dp-branes, interacting with the  d=10 supergravity background. In both theories we encounter the non-linear interacting goldstino fermion representing fermionic degrees of freedom on the world-volume of the super-Dp-branes interacting with NS-NS 2-forms as well as with all RR form fields. 

Particularly interesting examples of such models  presented in \cite{Kallosh:2014vja}, which suggest  the supersymmetric versions of the axion monodromy \cite{McAllister:2014mpa}, may be given as 
\be
W= S\Bigl[f(\Phi) + A\sin (\alpha \Phi)\Bigr], \qquad K=-{(\Phi- \bar\Phi)^2\over 2} + S\bar S - g (S\bar S)^2\ .
\ee
Here $\Phi =a + i \phi$.  For $S = \phi = 0$, one finds the  inflaton potential
\be
V = \Bigl[f (a ) + A\sin (\alpha  \, a)\Bigr]^2 \ .
\ee
The term $g (S\bar S)^2$ was  introduced in \cite{Kallosh:2014vja} for stabilization of the field $S$ at $S = 0$, and the inflaton $a$ is the combination $\Phi + \bar \Phi$ not appearing in the K\"ahler potential. 

To have a stringy interpretation of these models requires 
to take the superfield $S$ to be a nilpotent one, which is a valid step for these models. This removes the term $g (S\bar S)^2$ from the \K\, potential. In  new models with $S^2=0$ the term $g (S\bar S)^2$ is no longer required, and it also vanishes.

The \K\, potential for the inflaton multiplet in these models
\be
K=-{(\Phi- \bar\Phi)^2\over 2} 
\label{Kinfl}\ee
still requires a string theory interpretation.  In case of  Calabi-Yau type compactification,  which leads to N=2 special geometry,  one would expect  \K\, potentials of the logarithmic form with shift symmetry 
\be
K= c \ln [(z_0-\bar z_0)^2 - (z_i-\bar z_i)^2] \ .
\ee
It was suggested in \cite{Kallosh:2007ig} that in such case, if the modulus $z_0$ is stabilized, one might expand such a logarithm. If we keep just one of the field $z_i$, we find the expression 
\be
K=  c' \ln [1 - (z'_1-\bar z'_1)^2] \approx -c' [(z'_1- \bar z'_1)^2 ]
\ee  
for the inflaton \K\, potential \rf{Kinfl} of the desired type.

The remaining steps require to find specific string theory models and a choice of the form-field and a super-Dp-brane which would lead to a more specific choices of the superpotentials. But here, again, we remind that once the interaction between the fermion goldstino and any d=10 supergravity field related to $\Phi$ is established, in d=4 supergravity we can only use the superpotential $W=Sf(\Phi)$ since there is nothing else available due to $S^2=0$ condition. Terms independent on $S$ do not have this interaction whereas all higher powers of $S$ starting with $S^2$,  vanish.

To summarize, the new model, a candidate for an axion monodromy in string theory, has a d=4 bosonic supergravity action with one nilpotent superfield
\be
W= S\Bigl[f(\Phi) + A\sin (\alpha \Phi)\Bigr],\quad K=-{(\Phi- \bar\Phi)^2\over 2} + S\bar S \qquad {\rm at} \qquad S^2=0 \ .
\ee

Another example is given by the $\alpha$-attractor model \cite{Cecotti:2014ipa}, \cite{Kallosh:2014rga}. Here  we just present a new construction with a nilpotent multiplet $C$
\be
K= -3\, \alpha \log \left(T + \bar T   -  C \bar C  
 \right)\, , \qquad 
 W=  C F(T) \qquad {\rm at} \qquad C^2=0 \ .
\label{sugra}\ee
In the special case of this model at $\alpha=1$ and $F(T)= a+b T$, which leads to  Starobinsky model of inflation, it was  was shown in \cite{Antoniadis:2014oya} how to switch from the unconstrained $C$ to a nilpotent one. Here we explain it for generic $\alpha$ and generic functions $F(T)$.
The bosonic part of the supergravity model  at  $C|_{\theta=0}=c=0$  is given by the following expression 
\be
e^{-1} {\cal L}|_{c=0} = {1\over 2 } R - 3\alpha {\partial T \partial \bar T\over (T+\bar T)^2}- {1\over 3} {F(T) F(\bar T)\over (T+\bar T)^{3 \alpha -1} } \ .
\ee
When the imaginary part of the $T$-field is stabilized, the action becomes at $T=\bar T=t$ and $c=0$
\be
e^{-1} {\cal L} = {1\over 2 } R - {3\over 4} \alpha \Big ({\partial  t\over  t} \Big )^2- {1\over 12} \tilde f^2 (t)   \,.
\ee
Here $\tilde f(t)= F(t) t^{(1-3\alpha)/2}$.
In canonical variables $
T= e^{\sqrt{2\over 3\alpha}\vp}$ and using the fact that 
$
 \tilde f \Big (e^{\sqrt {2\over 3\alpha}\,  \vp}\Big )= f\Big(\tanh {\varphi\over\sqrt{6\alpha}}\Big)
$ one find the action for the inflaton in the form
\be
e^{-1} {\cal L} = {1\over 2 } R - {1\over 2}  (\partial \vp)^2 - f^2\Big(\tanh {\varphi\over\sqrt{6\alpha}}\Big) \ .
\ee
In this form one can recognize it as  conformal inflation  universality 
class attractor models  in \cite{Kallosh:2013pby}. Above we described a supersymmetric embedding of this class of models, following  \cite{Cecotti:2014ipa}, \cite{Kallosh:2014rga}.
In \rf{sugra}  we actually present  a simpler  version of  supersymmetric models in \cite{Cecotti:2014ipa}, \cite{Kallosh:2014rga} since we are now using  a nilpotent superfield $S$.

We have two comments on these models. First, at $\alpha=1$ we can add to the superpotential a constant term
\be
K= -3\,  \log \left(T + \bar T   -  C \bar C \right)\, , \qquad W=  C F(T) + W_0 \qquad {\rm at} \qquad C^2=0 \ .
\label{sugraY}\ee
As always in the no-scale case, this will not affect anything in our bosonic model, the potential will be the same as above, however, the fermionic action will be different, for example the gravitino will have a contribution to the mass term due to $W_0$.

Our second comment is about the choice of $\alpha$ from the string theory perspective. One would expect that $3\alpha=n$ where $n=1, 2, 3$ which means that $\alpha= 1/3, 2/3, 1$. 
\be
K= -n \log \left(T + \bar T   -  C \bar C  
 \right)\, , \qquad 
 W=  C F(T)  \qquad {\rm at} \qquad C^2=0 \ .
\label{sugraX}\ee
In view of the fact that for all these attractor models with generic $F(T)$ the prediction for gravity waves $T/S=r$ depends on $n$ as $r= {12 \alpha\over N^2}$  \cite{Kallosh:2014rga}  we find now that
\be
r={4n\over N^2} \ ,
\ee 
where $N$ is a number of e-foldings. In this form the inflationary attractor model has a simple relation to  D-brane actions. 

The new manifestly supersymmetric 
 superconformal action for the $\alpha$-attractor models is 
\begin{align}\label{alpha1}
-\bigg[ \bar X^0 X^0 \left(T + \bar T   -  C \bar C  
 \right)^ {\!\!\alpha}\, \bigg]_D+\Big(\big[C F(T)(X^0)^3+ \Lambda ({X^0})^2 C^2\big]_F  + h.c.\Big).
\end{align}
where the nilpotency of the superfield $C$ is imposed as a result of a solution of the equation of motion over the Lagrange multiplier superfield $\Lambda$. It differs from the related action in \cite{Cecotti:2014ipa} by the absence of the stabilization term depending on $( C \bar C)^2$ and by the presence of the term $\Lambda ({X^0})^2 C^2$.
And   now the action \rf{alpha1} is associated with the D-brane actions.

\subsection{Exit from inflation }

The general class of models which we study in this paper has a potential $V= e^{K(\vp)} K^{S\bar S} |f(\varphi)|^2$, where $\varphi$ is the inflaton, either $\phi$ or $a$, defined earlier. At the minimum of the potential with $f(\phi) = f_{0}$, there are two possibilities, one is that $f_{0}\neq 0$ and the other is that $f_{0}=0$.
If  $f_{0}\neq 0$, the potential is positive at the minimum, $V>0$.
If $f_{0}=0$ , the potential also vanishes at the minimum, $V=0$. In the bosonic theory there is no significant difference between these two cases. However, in our new models with the nilpotent multiplet, the fermionic sector of the theory is highly sensitive to this difference: there are many terms in the fermionic action which have negative powers of $f$, see for example \rf{VA}, or the super-Dp-brane action in \rf{actionBI} where $\alpha=f^{-1}$.

In models where  $f(\varphi)$ does not vanish at the minimum, the exit of inflation takes place in  de Sitter space and the fermionic action at the minimum of the potential is well defined since $f_{0}\neq 0$. 
If, however, $f_{0}=0$, the fermionic part of the action appears to become singular. However, the careful procedure of taking the limit to $f\rightarrow 0$ in the  action of the D-brane involves a redefinition of the fields 
\be
\lambda_\alpha = f \tilde \lambda_\alpha \ ,
\ee
and the same for vectors, $F_{\mu\nu}= f \tilde F_{\mu\nu}$, if they are present. Replacing also $\alpha^{-1}$ by $f$ in the DBI-VA  action \rf{actionBI} we find
\begin{equation}
\label{actionBInew} S_{DBI-VA} =  - f^2 \int \rmd^{10} \sigma\,\left\{
\sqrt{- \det (\tilde G_{\mu\nu} +  \tilde {\cal F}_{\mu\nu})}  \right\}\,,
\end{equation}
\begin{eqnarray}
\tilde G_{\mu\nu} = \eta_{mn} \tilde \Pi_\mu^m \tilde \Pi_\nu^n  \qquad  \Pi_\mu^{m} &=& \delta^{m}_{\mu}
-\bar{\tilde \lambda} \Gamma^{m}
\partial_\mu \tilde \lambda \ , \end{eqnarray}
\begin{eqnarray}
\tilde {\cal F}_{\mu\nu} &\equiv& \tilde F_{\mu\nu} - 2\bar{\tilde {\lambda}}\Gamma_{[\nu }\partial_{\mu]}\tilde \lambda\, .
\end{eqnarray}
When the limit $f\rightarrow 0$ in the  action of the D-brane is taken with fields $\tilde \lambda_\alpha$ and $\tilde F_{\mu\nu}$  fixed, the total action of the D-brane vanishes. This is consistent with the fact that the total $S$ multiplet disappears. During inflation when $f>0$ the fermionic goldstinos exist in the action in agreement with the nilpotent $S^2=0$ multiplet, however, when $f\rightarrow 0$ it means that 
$\lambda \lambda /f $ becomes $f \tilde \lambda \tilde \lambda$ and disappears in the limit $f\rightarrow 0$. In such case the degrees of freedom on the D-brane decouple near the exit from inflation.

\section{Manifestly Supersymmetric Uplifting Using  Dp-branes}

Adding  other fields  and taking more general  superpotential $W$ by adding an $S$-independent part,
\be
W= S f(\Phi, T^i) + W(T^i) \ ,
\label{gen}\ee
 where $S$ is a nilpotent field, may allow us to uplift AdS and Minkowski vacua to dS as well as to  study more general inflationary models.
 Even more general models of cosmology may be studied, which have more chiral nilpotent superfields, as well as other unconstrained chiral superfields. According to the superconformal action \rf{confConstrA}, supergravity models with any number of chiral multiplets and nilpotent chiral multiplets are now available. We expect that these models will be studied in the future.

 A combination of models including string theory volume modulus used in the KKLT models \cite{Kachru:2003aw} or  KL models \cite{Kallosh:2004yh} with some other superfields, matter multiplets and hidden sector superfields including the so-called Polonyi models \cite{Polo}, was constructed and studied in \cite{Lebedev:2006qq,Dine:2006ii} and \cite{Kallosh:2006dv}. In all of these models, the superfield $S$ is an unconstrained superfield, which is either zero  or takes some other constant value at the minimum of the potential. Its presence in the theory helped to uplift AdS or Minkowski vacua of the KKLT-type models to dS vacua. However, as we already mentioned, it was not easy to find an interpretation of the superfield $S$ from the string theory perspective.

 In this section, we will give a brief overview of the new approach to uplifting when the superfield $S$ in \rf{gen} is nilpotent.

\subsection{O'KKLT uplifting with the nilpotent multiplet}

To explain how things changed now when we restrict $S$ by the nilpotent condition, $S^2=0$, let us look at the O'KKLT models in \cite{Kallosh:2006dv}, where O' refers to the underlying O'Raifeartaigh model. In this model there are two relatively heavy fields which are integrated out. This leads to the   effective O'KKLT supergravity models with                                                                                                                                                                                                                                                                                                                                                                                                         \be
W=W_0 + Ae^{-a\rho}- \mu^2 S, \qquad K= - 3 \ln(\rho + \overline{\rho})+ S\bar S - {(S\bar S)^2\over \Lambda^2} \ .
\label{model1a}\ee
The complete potential $V(\sigma, \alpha, x, y)$ as a function of 4 scalars,
\be
\rho= \sigma+i\alpha \ , \qquad S=x+iy \ .
\ee
at  small $S\bar S$,  can be represented  in a rather compact form
 \be
V_{O'KKLT}= V_{KKLT}(\rho, \bar \rho) + {V_{O'}(S, \bar S) \over (\rho+\bar \rho)^3} - i(S-\bar S) V_3  + (S+ \bar S) V_4 +  S\bar S V_5  \ .
\label{potential}\ee
Here the potential of the quantum corrected O'Raifeartaigh model $V_{O'}(S, \bar S)$ is
\be
V_{O'}(S, \bar S) = \mu^4 e^{{S\bar S(\Lambda^2-S\bar S)\over \Lambda^2}} \left [{\left(\Lambda^2(1+(S\bar S)-2(S\bar S)^2\right)^2\over \Lambda^4- 4 \Lambda^2 S\bar S }- 3S\bar S\right] \ ,
\ee 
and separately is has a minimum at $S=x+iy=0$.
$V_3 (\rho, \bar \rho, S, \bar S) $,  $ V_4 (\rho, \bar \rho, S, \bar S)$ and   $ V_5 (\rho, \bar \rho, S, \bar S)$ depend on $S$, $\bar S$  polynomially.

The KKLT potential $V_{KKLT}(\rho, \bar \rho)$, taken separately, has an AdS minimum at the vanishing axion,  $\alpha = 0$,  and at some (large) value of $\sigma$.  It was established in \cite{Kallosh:2006dv} that the values of the axion fields $\alpha$ and $y$ at the minimum of the combined potential remain equal to zero, whereas the values of $\sigma$ and $x$ are  slightly shifted.

\noindent According to the rules explained earlier, the potential in the new O'KKLT model follows from
\be
W=W_0 + Ae^{-a\rho}- \mu^2 S\, , \qquad K= - 3 \ln(\rho + \overline{\rho})+ S\bar S\,  \qquad {\rm at} \qquad S^2=0 \ .
\label{modelN}\ee
Here $S$ in the nilpotent generalization of the Polonyi field. Now,  after computing the potential,  we have to set the scalar part of the superfield $S$ to zero. Therefore we do not need the stabilization term $- {(S\bar S)^2\over \Lambda^2}$.
We find  
\be
V_{New\, O'KKLT}= V_{KKLT}(\rho, \bar \rho) + {\mu^4 \over (\rho+\bar \rho)^3}   \ .
\label{potentialNew}\ee
This shows that  \rf{modelN} corresponds to a manifestly supersymmetric version of  uplifting  of the KKLT model (improving the purely bosonic expression for the uplifting term from the anti-D-3 brane used in \cite{Kachru:2003aw}).

In case we would start with the model 
\be
W=W_0 + Ae^{-a\rho}- \mu^2 S\, , \qquad  K= - 3 \ln(\rho + \overline{\rho}-  S\bar S)\,  \qquad {\rm at} \qquad S^2=0  \ .
\label{modelN2}\ee
the uplifted potential would be
\be
V_{New\, O'KKLT}^{warped}= V_{KKLT}(\rho, \bar \rho) + {\mu^4 \over (\rho+\bar \rho)^2}   \ .
\label{potentialNew1}\ee
as expected in the situation with warping \cite{Kachru:2003sx}. Here one can see it from our general formula $V_{eff}= e^K K^{S\bar S} |W_S|^2$.

Thus we have shown here that once the uplifting O'KKLT-type models used in \cite{Lebedev:2006qq,Dine:2006ii,
Kallosh:2006dv} are modified to include a nilpotent chiral multiplet, they become string theory motivated via Dp-branes and provide a manifestly supersymmetric uplifting to dS vacua for numerous AdS vacua in the stringy landscape. The price for this is a non-linearly  realized spontaneously broken supersymmetry of the Volkov-Akulov type with a complicated fermion action,  which is  present on the world-volume of the Dp-branes. 

\subsection{More models with Polonyi superfield replaced by a nilpotent one}

A similar generalization/simplification is available for the recent string theory motivated analytic classes of metastable de Sitter vacua where only the unconstrained chiral superfields are involved \cite{Kallosh:2014oja}.
One may start with the KL model \cite{Kallosh:2004yh} with $K=  -3 \log (T + \bar T)$ and the racetrack potential
\be
W_{\rm KL}(T) = W_0 + Ae^{-aT}- Be^{-bT} \ .
\label{adssuprace}
\ee
The term $Be^{-bT}$ allows the new model to have a supersymmetric Minkowski solution. Indeed,
for the particular choice of $W_0$,
\be\label{w0}
W_0= -A \left({a\,A\over
b\,B}\right)^{a\over b-a} +B \left ({a\,A\over b\,B}\right) ^{b\over b-a} ,
\ee
the potential of the field $T$ has a supersymmetric minimum $T_{0}= {1\over a-b}\ln \left ({a\,A\over b\,B}\right)$ with
$W_{\rm KL} (T_{0})=0$,  $D_\rho W_{\rm KL}(T_{0}) = 0$,   and $V(T_{0})=0$.
To achieve supersymmetry breaking  one can add to this model the Polonyi field $C$.
The K\"ahler and superpotential are
\be\label{pol}
K = K(T) + C \bar C - \frac{(C \bar C)^{2}}{\Lambda^2}\ , \qquad W = W(T) +\mu_{1}+ \mu_2 C   \ .
\ee
Here $\mu_{i}$ are supposed to be very small.  Depending on the relation between $\mu_{i}$, this may either lead to a downshift of the Minkowski minimum, making it AdS (for $\mu_{2}^{2}< 3 \mu_{1}^{2}$), or uplift it to a dS minimum (for $\mu_{2}^{2}> 3 \mu_{1}^{2}$). To obtain a slightly uplifted state with the present value of the cosmological constant $ \sim 10^{{-120}}$, one should have $\mu_{2}^{2}\approx 3 \mu_{1}^{2}$.
In this case  $m_{C}^{2} =  {3\mu_{1}^{2} \over 2T_{0}^{3} \Lambda^{2}} $, which becomes superheavy in the limit $\Lambda\to 0$ \cite{Dudas:2012wi, Kallosh:2014oja}.

What happens to this scenario if one takes the Polonyi field $C$ which belongs to the nilpotent multiplet? This field vanishes, which is similar to what happens in the model considered above in the limit $\Lambda \to 0$. However, now we do not need the stabilization term $ \frac{(C \bar C)^{2}}{\Lambda^2}$, and we have string theory interpretation of the uplifting. 

Moreover, in this scenario the Polonyi field $C$ does not cause the famous  cosmological moduli problem, which bothered cosmologists for more than three decades \cite{Polonyiproblem}. This problem does not appear  because this superfield is nilpotent, and therefore the scalar  vanishes by construction.

The situation with uplifting in other string theory models is very similar.
One of the examples is the STU model with a Minkowski vacuum with all moduli stabilized, with
\begin{equation}\label{eq:K}
K(S,T,U) = -\log (S + \bar S) -3 \log (T + \bar T)-3 \log (U + \bar U) \,,
\end{equation}
\be\label{simplemin}
W(S,T,U) = A\,(S - S_{0}) (1 - c\, e^{-a\,T}) + B\,(U - U_{0})^2 \ .
\ee
The potential has a stable supersymmetric minimum at $S = S_{0}$, $U = U_{0}$ and $T = {\log c\over a}$. Just as in the KL model, one can uplift this stable Minkowski vacuum to a metastable dS vacuum by adding the Polonyi field $C$ as we did in (\ref{pol}) with $\mu_{1} \sim \mu_{2} \ll 1$, and $\Lambda \ll 1$ \cite{Kallosh:2014oja}:
\be\label{pol2}
K = K(S,T,U) + C \bar C - \frac{(C \bar C)^{2}}{\Lambda^2}\ , \qquad W = W(S,T,U) +\mu_{1}+ \mu_2 C\ .
\ee
In fact, for some parameters of this model, uplifting can be realized even in the absence of the stabilizing term $- \frac{(C \bar C)^{2}}{\Lambda^2}$; however, this term certainly helps.

Once again, for the nilpotent Polonyi field $C$, we do not need any stabilization terms. The field $C$ vanish as in the original model in the limit $\Lambda ->0$.  Thus, the uplifting, which was realized in the original model in \cite{Kallosh:2014oja}, is achieved even easier in the model with the nilpotent Polonyi field $C$,
\be\label{pol3}
K = K(S,T,U) + C \bar C \ , \qquad W = W(S,T,U) +\mu_{1}+ \mu_2 C \qquad {\rm at} \qquad C^2=0
\ .
\ee

\section{Discussion}
Volkov-Akulov construction of a non-linearly realized supersymmetry \cite{Volkov:1973ix} had a purpose of describing a massless Goldstone spin 1/2 fermion in Minkowski space, for example neutrino. This supermultiplet does not have a scalar spin 0  partner, as different from the models with linear supersymmetry. VA theory was invented before we knew that neutrino is not massless and the space-time is not Minkowski but de Sitter with a cosmological constant $\Lambda\sim 10^{-120} M_P^4$. Several authors noted that this parameter could be related to the neutrino mass as $m_{\nu}\sim \Lambda^{1/4} \sim 10^{-2} $eV. This relation remains puzzling and suggests that, perhaps, better understanding of the Volkov-Akulov construction and developing on it might be useful. In particular, the general investigation of the fermionic sector of this theory is in order, if one would like to relate it to particle phenomenology.

In our paper, we concentrated on other aspects of the VA construction, related to its bosonic sector. We studied general cosmological issues such as inflation and string theory moduli stabilization by including the  VA supermultiplet interacting with supergravity and other chiral superfields. The technical tool for including Volkov-Akulov supermultiplet was to use it in the form of a nilpotent chiral multiplet, $S^2=0$, as suggested  by Rocek in  \cite{rocek}. The first important modification of the supersymmetric version of the Starobinsky inflation \cite{Cecotti:1987sa}, \cite{Kallosh:2013lkr} by replacing one of the superfields in this model by a nilpotent one was  made in \cite{Antoniadis:2014oya}.

 In this paper we found that a large number of the  previously studied inflationary models in supergravity \cite{Cecotti:1987sa,Kallosh:2013lkr,Kawasaki:2000yn,Kallosh:2010xz,Kallosh:2014vja,Kallosh:2014xwa,Kallosh:2013pby,Cecotti:2014ipa,Kallosh:2014rga} can be easily updated to replace one the superfields by the nilpotent one.
The superpotential of these models is linear in a chiral superfield $S$, which in the new versions of these models  has to be replaced by the nilpotent $S$ satisfying the constraint $S^2=0$, so that these new models now include the VA goldstino supermultiplet. All these models
are significantly simplified when one of the superfields is nilpotent, and we explained the relation between the old and new models. The bosonic part of the theory is simpler, and since only the bosonic part is immediately relevant to inflationary cosmology, the new inflationary models look significantly more attractive. The scalar component of the nilpotent supermultiplet is replaced by the fermion bilinear, it does not need to be stabilized, many terms in the bosonic action vanish, investigation of the existing inflationary models is considerably simplified, and many new inflationary models become possible.

Another interesting aspect of the new model is that the Dp-brane actions in string theory are ultimately related to the VA actions \cite{Kallosh:1997aw,Bergshoeff:2013pia}.  The fermions which leave on the world-volume of the Dp-brane have a non-linearly realized spontaneously broken supersymmetry. In this sense, our new models of inflation with one of the superfields replaced by a nilpotent one, originate from string theory. This means, in particular, that this set of models may provide a manifestly supersymmetric basis for axion monodromy supergravity models related to string theory via Dp-branes interacting with the supergravity background. We also demonstrated that the new models with the nilpotent goldstino multiplet provide a simple manifestly supersymmetric uplifting mechanism in the KKLT-type constructions.

\section*{Acknowledgments}

We are grateful to A. Antoniadis, E. Bergshoeff,  C. Cecotti, E. Dudas, K. Olive,   A. Van Proeyen and  A. Westphal for  collaboration on closely related recent projects, on which our current work is based. We would like to thank S. Kachru, M. Porrati,   D. Roest, A. Sagnotti,  E. Silverstein, B. Vercnocke and T. Wrase for enlightening conversations. 
SF is supported by ERC Advanced Investigator Grant n. 226455 Supersymmetry, Quantum Gravity and Gauge Fields (Superfields). RK and AL are supported by the SITP and by the NSF Grant PHY-1316699 and RK is also supported by the Templeton foundation grant `Quantum Gravity Frontiers'.

\end{document}